\documentclass[11pt]{article}
\usepackage{amsmath}
\DeclareMathOperator*{\plim}{plim}
\usepackage{subfig}
\usepackage{color}
\usepackage{multirow}
\usepackage{natbib}
\usepackage{graphicx}
\usepackage[dvips]{epsfig}
\usepackage{lingmacros}

\newcommand{\be}{\begin{eqnarray}}
\newcommand{\ee}{\end{eqnarray}}
\newcommand{\ba}{\begin{eqnarray*}}
\newcommand{\ea}{\end{eqnarray*}}
\newtheorem{theorem0}{Theorem}
\newtheorem{lemma0}{Lemma}

\newenvironment{lemma}{\begin{lemma0}}{\end{lemma0}}
\newenvironment{proof}{\noindent{\bf Proof.} \rm}{ \hfill \fbox{}}
\def\boldfacefake #1{%
    \hbox{%
        \mathsurround=0pt
        \hbox to 0.4pt{$#1$\hss}%
        \hbox to 0.4pt{$#1$\hss}%
        \hbox {$#1$}%
    }%
}
\newcommand{\nmathbf}{\bf}
\def\bfI{\mbox{$\nmathbf I$}}
\def\bfJ{\mbox{$\nmathbf J$}}
\def\bfx{\mbox{$\nmathbf x$}}
\def\bfY{\mbox{$\nmathbf Y$}}
\def\bfy{\mbox{$\nmathbf y$}}
\def\bfSigma{\boldfacefake{\Sigma}}
\def\bfPsi{\boldfacefake{\Psi}}

\def\bftheta{\boldfacefake{\theta}}

\newcommand{\iid}{\stackrel{\mathrm{\textsl{iid}}}{\sim}}
\newcommand{\go}{\rightarrow}
\newcommand{\vareps}{\varepsilon}

\newcommand{\qed}{\nobreak \ifvmode \relax \else
      \ifdim\lastskip<1.5em \hskip-\lastskip
      \hskip1.5em plus0em minus0.5em \fi \nobreak
      \vrule height0.75em width0.5em depth0.25em\fi}

\begin{document}
\title{\bf Bayes Factor Consistency for One-way Random Effects Model}
\author{Min Wang \ \ and \ \ Xiaoqian Sun\\
 {\small
  Department of Mathematical Sciences, Clemson University, Clemson,
   SC 29634, USA}
   }
\date{}        % Enter your date or \today between curly braces
\maketitle

\begin{abstract}

In this paper, we consider Bayesian hypothesis testing for the balanced one-way random effects model. A special choice of the prior formulation for the ratio of variance components is shown to yield an explicit closed-form Bayes factor without integral representation. Furthermore, we study the consistency issue of the resulting Bayes factor under three asymptotic scenarios: either the number of units goes to infinity, the number of observations per unit goes to infinity, or both go to infinity. Finally, the behavior of the proposed approach is illustrated by simulation studies.


\end{abstract}

\section{Introduction} \label{section1}

Consider the balanced one-way analysis-of-variance (ANOVA) random effects model
\be \label{model1}
y_{ij} = \mu + a_i +  \vareps_{ij} \quad \mbox{for} \ \ i = 1, 2, \dots, p, ~j = 1, 2, \ldots, r,
\ee
where $y_{ij}$ is the $j$th observation associated with the unit $i$ and $\mu$ represents the unknown intercept. Here $p$ $(\geq 2)$ is the number of units and $r$ $(\geq 2)$ is the number of observations per unit. It is assumed that the random effect $a_i$ and the error term $\vareps_{ij}$ are mutually independent, and that $a_i \iid N(0, \sigma_a^2)$ and $\vareps_{ij} \iid N(0, ~\sigma^2)$ for all $i$ and $j$, where \textsl{iid} represents ``independent and identically distributed.'' The unknown parameters $\sigma^2_a$ and $\sigma^2$ are often called variance components in the literature. For notational convenience, throughout the paper, let $\sum_i$ and $\sum_j$ stand for $\sum_{i=1}^p$ and $\sum_{j=1}^r$, respectively.

In such a balanced variance components model (\ref{model1}), we are often interested in evaluating whether the random effects should be included, which is equivalent to testing
\be \label{hyp:1}
\label{test:org} M_0: \sigma_a^2 = 0 \quad \mbox{versus} \quad M_1: \sigma_a^2 \neq 0.
\ee

For ease of exposition, let $N_r (\bfx \mid \bftheta, ~\bfPsi)$ represent the probability density of a multivariate normal distribution with an $r$-dimensional random vector of observations $\bfx$, an $r$-dimensional mean vector $\bftheta$ and an $r \times r$ variance-covariance matrix $\bfPsi$. Without loss of generality, the model (\ref{model1}) can then be expressed compactly in matrix form as follows
\be \label{joint:density}
{f({\bfY} \mid \mu, ~\sigma^2, ~\sigma^2_a)} = \prod_{i=1}^{p}{N_r({{\bfy}_i} \mid \mu{\bf1}_r, ~\bfSigma)},
\ee
by letting $\bfY= (\bfy_1', \cdots, \bfy_p')'$ with $\bfy_i = (y_{i1}, \cdots, y_{ir})'$, where ${\bf{1}}_r$ is an $r \times 1$ vector of ones and $\bfSigma = \sigma^2 \bfI_r + \sigma^2_a \bfJ_r$ with $\bfI_r$ being an $r \times r$ identity matrix and $\bfJ_r$ being an $r\times r$ matrix containing only ones. Accordingly, the hypothesis testing problem (\ref{hyp:1}) can be equivalently expressed as testing the following two models
\be \label{hpy:0}
M_0: f_1(\bfY \mid \mu, \sigma^2, \sigma^2_a) = f(\bfY \mid \mu, \sigma^2, 0) \quad \mbox{versus} \quad M_1: f_2(\bfY \mid \mu, \sigma^2, \sigma^2_a) = f(\bfY \mid \mu, \sigma^2, \sigma^2_a).
\ee

It is well known by \cite{Box:Tiao:1973} that the classical unbiased estimates of $\sigma_a^2$ can be negative even if the true value of $\sigma^2_a$ is nonnegative. This is a serious disadvantage of using these estimates in frequentist analysis. To avoid this problem, this paper deals with the problem of hypothesis testing or model selection based on the Bayesian approach. As mentioned by one referee, an operational advantage of the Bayesian approach is that likelihood-based methods require special care since the parameter being tested is a boundary case, leading to the failure of standard asymptotic scenarios; see, for example, \cite{Mall:Zhou:1996}, \cite{Paul:Wake:Kass:1999} and references therein. In addition, there are many other advantages for using the Bayesian approach to this problem over the frequentist or classical approach. We here refer the interested reader to \cite{Westfall:1996} and \cite{Berg:Peri:2001} for more details.

From the Bayesian viewpoint, the Bayes factor (\citeauthor{Kass:95}, \citeyear{Kass:95}) offers a natural way of measuring the evidence in data for various competing models in terms of their posterior model probabilities. In our problem, the Bayes factor for comparing $M_1$ to $M_0$ given by (\ref{hpy:0}) can be written as
\be \label{m01:BF}
BF_{10} = \frac{m_{1}(\bfY)}{m_{0}(\bfY)},
\ee
where
\begin{equation*} \label{lik:1}
m_1(\bfY) = p(\bfY\mid M_1)  = \int\!\!\int\!\!\int{f_2(\bfY\mid\mu, \sigma^2, \sigma^2_a)\pi_1(\mu, \sigma^2, \sigma^2_a)}\, d\mu\, d\sigma^2\, d\sigma^2_a,
\end{equation*}
and
\begin{equation*}\label{lik:0}
m_0(\bfY) = p(\bfY\mid M_0) = \int\!\!\int{f_1(\bfY\mid\mu, \sigma^2)\pi_0(\mu, \sigma^2)}\, d\mu\, d\sigma^2
\end{equation*}
with $\pi_1(\mu, \sigma^2, \sigma^2_a)$ and $\pi_0(\mu, \sigma^2)$ being the joint prior densities for the unknown parameters under $M_1$ and $M_0$, respectively. From Bayes theorem, the posterior probability of model $M_1$ given $\bfY$ can be expressed through the Bayes factor as
\be \label{post:prob}
p(M_1\mid \bfY) = \frac{p(M_1)m_1(\bfY)}{p(M_1)m_1(\bfY) + p(M_0)m_0(\bfY)} =  \frac{p(M_1)BF_{10}}{p(M_0) + p(M_1)BF_{10}},
\ee
where $p(M_i)$ is the prior probability of model $M_i$ for $i=0,1$. In the absence of prior knowledge, it is natural to specify $p(M_0) = p(M_1) = 1/2$. Therefore, for decision-making, the model $M_1$ is more likely to be selected if $p(M_1\mid \bfY)>1/2$, or equivalently, $BF_{10} > 1$.

In the Bayesian framework, it is of particular interest to study the consistency issue of the proposed procedures. Here, consistency means that the true model will be selected if enough data are provided, assuming that one of the competing models is true. Let $M_T$ stand for the true model. \cite{Fern:2001} formally defined the posterior consistency for hypothesis testing or model selection as
\begin{equation}\label{consistency_def:1}
 \plim_{n \rightarrow \infty}p(M_T\mid Y) = 1,
\end{equation}
where `$\plim$' denotes convergence in probability as $n$, the total number of observations, goes to infinity. Due to the relationship between the posterior probability and the Bayes factor, the expression (\ref{consistency_def:1}) for our testing problem (\ref{hyp:1}) becomes
\begin{equation}
\plim_{n \go \infty}BF_{10}= \infty,
\end{equation}
if $M_1$ is the true model, whereas
\begin{equation}
\plim_{n \go \infty}BF_{10} = 0,
\end{equation}
if $M_0$ is the true model. Since $n=pr$ in this paper, we shall mainly focus on the consistency of Bayes factor for the hypothesis testing problem in the balanced one-way random effects model under three asymptotic scenarios as follows:

\noindent {\bf Scenario 1} \ \ $r$ goes to infinity, but $p \geq 2$ is fixed.

\noindent {\bf Scenario 2} \ \ $p$ goes to infinity, but $r \geq 2$ is fixed.

\noindent {\bf Scenario 3} \ \ both $r$ and $p$ go to infinity.

For the hypothesis testing problem in the balanced random effects model, numerous Bayesian approaches have recently been proposed in the literature. For example, \cite{Westfall:1996} proposed a new Bayes factor for hypothesis testing in the one-way ANOVA model with either fixed or random effects and then studied the Bayes factor consistency under the first two asymptotic scenarios. Later, \cite{GarcSun:2007} developed the divergence-based prior and the intrinsic prior for the parameter $\sigma_a^2$ and also showed that both priors produce consistent Bayes factors. In addition, \cite{cano:Kess:2007} derived a new Bayes factor based on the methodology of integral priors introduced by \cite{Cano:Salm:Robe:2008}. An attractive feature of the integral priors is that they take advantage of Markov chain Monte Carlo (MCMC) techniques to produce unique Bayes factors often, whereas it is unclear whether or not the integral priors for this testing problem are unique from the theoretical viewpoint. Additionally, they do not further investigate the consistency issue of the resulting Bayes factor.

It is worthwhile mentioning that the approaches mentioned above have been shown to perform well in a variety of real applications. In most cases, the integral representation is involved in the expression of Bayes factors, so numerical approximations will generally be employed. However, it is not an easy task in applied statistics to decide which type of approximations to be more appropriate, especially when both $p$ and $r$ are extremely large. Moreover, they do not seem to take the asymptotic property of the proposed testing procedures into account under Scenario 3, which would also be of interest to readers and researchers.

In this paper, we propose an explicit closed-form Bayes factor without integral representation for the balanced one-way random effects model. Of particular note is that the proposed Bayes factor is exactly the same as the one derived by \cite{Maruyama:2009} for the balanced fixed effects model. In addition, we study the consistency issue of Bayes factor under the three asymptotic scenarios mentioned above. It is shown that the resulting Bayes factor is always consistent under $M_0$, but it may be inconsistent under $M_1$ in Scenario 2 due to the presence of a small inconsistency region, which can be characterized by the number of observations per unit.

One may argue that making the distinction between fixed effects and random effects is obscure from the Bayesian viewpoint because all parameters could be viewed as random variables. Nevertheless, as mentioned by one referee, one main difference between the two models is that for the random effects model, the dimension of the parameter space under the full model is three, namely $(\mu, \sigma_a, \sigma)$, which does not grow when either the number of observations approaches infinity or the number of units approaches infinity, whereas for the fixed effects model, the model dimension grows as the number of units increases. In addition, there are two main differences between the present paper and the study of \cite{Maruyama:2009} for the fixed effects model.
\begin{itemize}
\item [(i)] In the random effects model, under orthogonality and same magnitudes, one can easily justify the reasonability of using the same (even noninformative) prior for the common parameters $\mu$ and $\sigma^2$. We here refer the reader to \cite{GarcSun:2007} for more details. Specifically, we consider the consistency property of Bayes factor under a scenario in which both $r$ and $p$ approach infinity.
\item [(ii)]  From the Bayesian standpoint, both fixed effects model and random effects model can be treated as three-stage hierarchical models. As stated by \cite{Smit:1973}, `` \textit{for the Bayesian model the distinction between fixed, random and mixed models reduces to the distinction between different prior assignments in the second and third stages of the hierarchy.}'' For a detailed discussion on the topic, one may also refer to \cite{Rend:2002}. It is noteworthy that the prior formulations for the unknown parameters in this paper are different from the priors adopted by \cite{Maruyama:2009}.
\end{itemize}

The remainder of this paper is organized as follows. In Section \ref{section2}, we discuss the prior formulation for the unknown parameters $\mu$, $\sigma^2$ and $\sigma_a^2$, and then adopt a specific prior distribution for the ratio of variance components $\sigma_a^2/\sigma^2$, which results in an explicit closed-form Bayes factor without integral representation. In Section \ref{section3}, we investigate the corresponding consistency of Bayes factor under the three different asymptotic scenarios listed above. The performance of the proposed Bayes factor is illustrated through several simulated studies in Section \ref{section4}. Some concluding remarks are given in Section \ref{section5}. Finally, several useful lemmas and proofs will be provided in Appendix.

\section{Bayes factor} \label{section2}

Direct use of improper priors is unsuitable for the hypothesis testing problem because it may yield a Bayes factor up to some undetermined normalizing constants. Intrinsic priors, developed by  \cite{Berg:Peri:1996}, have been widely used to overcome this potential difficulty on the use of improper priors. The idea of intrinsic priors is to convert improper priors into ones suitable for computing the Bayes factors. We do not review them here, but rather point the interested reader to \cite{Berg:Peri:1998}, \citeauthor{More:Bert:Fran:1999} (\citeyear{More:Bert:Fran:1999}, \citeyear{More:eiro:Javi}, \citeyear{More:Giro:2008}), \cite{Case:Geor:More:2006}, \cite{Giro:Mart:Mero:2006}, \cite{Case:Geor:Gir:2009}, \cite{Torr:More:Giro:2011}, among others.

An alternative way to avoid such a pitfall of the Bayes factor when using improper priors is to choose the same improper prior for ``common parameters'' that appear in the two competing models, although it could be argued that the common parameters may change meanings from one model to another. Nevertheless, as mentioned by \cite{Kass:Vaid:1992}, under orthogonality (i.e., the expected Fisher information matrix is diagonal), the Bayes factor is quite robust to the selection of the same (even noninformative) prior adopted for the common orthogonal parameters. For the hypothesis  testing problem (\ref{hpy:0}), \cite{GarcSun:2007} showed that the common parameters $\mu$ and $\sigma^2$ are approximately (for a moderate or large value of $n$) orthogonal to the new parameter $\sigma_a$ in $M_1$. As a result, both $\mu$ and $\sigma^2$ may be assumed to have the same meanings in both $M_0$ and $M_1$ (\citeauthor{Jeff:1961}, \citeyear{Jeff:1961}, Chapter 5), justifying the use of the same noninformative priors. Accordingly, under $M_0$, we can adopt the following prior for $\mu$ and $\sigma^2$ given by
\be \label{prior:1}
\pi_0 (\mu, \sigma^2) = \frac{c}{\sigma^2},
\ee
where $c$ is a constant, and under $M_1$, we specify
\be \label{prior:2}
\pi_1 (\mu, \sigma^2, \sigma^2_a) = \pi_0 (\mu, \sigma^2)\pi^{\ast}(\sigma^2_a \mid \sigma^2),
\ee
where a scale family prior for $\sigma^2_a$ is adopted and given by
\be  \label{prior:3}
\pi^{\ast} (\sigma_a^2 \mid \sigma^2) = \frac{1}{\sigma^2}\pi\biggl(\frac{\sigma^2_a}{\sigma^2}\biggr),
\ee
with $\pi(\sigma^2_a/\sigma^2)$ being the prior distribution for the ratio of variance components $\sigma^2_a/\sigma^2$, which will be specified later.

Note that the idea of using same noninformative priors for common (orthogonal) parameters has been proved to be successful by many statisticians; see, for example, \cite{GarcSun:2007}, \cite{Baya:Garc}, \cite{liang:2008}, to name just a few. To avoid the undefined Bayes factors, a proper prior distribution is often required for the ratio of variance components $\tau = \sigma^2_a/\sigma^2$. According to Proposition 1 of \cite{GarcSun:2007}, the Bayes factor given by (\ref{m01:BF}) for the priors (\ref{prior:1}) and (\ref{prior:2}) along with (\ref{prior:3}) can be written as
\be \label{lik:integral}
BF_{10} =  \int_0^\infty{(1 + \tau r)^{-(p-1)/2}\biggl(1 - \frac{\tau r}{1 + \tau r}\frac{W_H}{W_T}\biggr)^{-(n-1)/2}} \pi(\tau)\, d\tau,
\ee
where $W_H$ and $W_T$ stand for the sum of squares between groups and the total sum of squares, respectively, and they are given by
\begin{equation*}
W_H = r\sum_i(\bar{y}_{i\cdot} - \bar{y}_{\cdot\cdot})^2 \quad \mbox{and} \quad W_T = \sum_{i}\sum_{j}(y_{ij} - \bar{y}_{\cdot\cdot})^2,
\end{equation*}
with $\bar{y}_{\cdot\cdot} = \sum_{i}\sum_{j}{y_{ij}}/n$ and $\bar{y}_{i\cdot} = \sum_{j}{y_{ij}}/r$. Various choices of the prior distribution $\pi(\tau)$ for the ratio of $\tau$ have recently been proposed in the literature. For example, \cite{Westfall:1996} advocated the following prior distribution
\be \label{hyp:gprior}
\pi^{WG}(\tau) = (1 + \tau)^{-2}I_{(0, \infty)}{(\tau)},
\ee
which is also named the hyper-$g$ prior in \cite{liang:2008}. Furthermore, \cite{Westfall:1996} showed the consistency of Bayes factor with the choice of prior (\ref{hyp:gprior}) for $\tau$ under the first two asymptotic scenarios above. Later, \cite{GarcSun:2007} proposed the intrinsic prior and the divergence-based prior for $\tau$ and then investigated the corresponding consistency of Bayes factor under the two priors, respectively. As suggested by one referee, it deserves to mention here that the divergence-based prior, developed by \cite{Baya:Garc:2007}, is a density function proportional to a positive measure of divergence between two competing models raised to a negative power $q$. One may also refer to \cite{GarcSun:2007} in detail.

In this paper, we adopt a new prior density for $\tau$, often called the Pearson type VI distribution with shape parameters $\alpha > -1$, $\beta > -1$ and scale parameter $\kappa > 0$. The density function of this distribution is given by
\be \label{beta:prime}
\pi^{PT}(\tau) = \frac{\kappa(\kappa\tau)^\beta (1 +\kappa\tau)^{-\alpha-\beta-2}}{B(\alpha + 1, \beta + 1)}I_{(0, \infty)}{(\tau)},
\ee
where $B(\cdot, \cdot)$ is the beta function. Note that the beta-prime distribution used by \cite{Maruyama:2009} is just a special case of the Pearson type VI distribution with $\kappa = 1$, and that $\pi^{WG}(\tau)$ in (\ref{hyp:gprior}) is also a special case with $\kappa=1$ and $\alpha = \beta=0$. To obtain an explicit closed-form Bayes factor, cases for which $\kappa = r$ will be of interest to us in what follows. With the use of transformation $t = r\tau$, simple algebra shows that the Bayes factor in (\ref{lik:integral}) with $\pi(\tau)$ replaced by $\pi^{PT}(\tau)$ in (\ref{beta:prime}) becomes
\begin{align} \label{BF:2F1}
BF_{10} &= \frac{1}{B(\alpha + 1, \beta + 1)}\int_0^\infty t^\beta \big(1 +t\big)^{(n-p)/2-\alpha-\beta-2}\biggl(1+ \frac{W_E}{W_T} t\biggr)^{-(n-1)/2}\, dt,
\end{align}
where $W_E$ represents the sum of squares within groups and is given by
\begin{equation*}
W_E = W_T - W_H = \sum_i\sum_j(y_{ij} - \bar{y}_{i\cdot})^2.
\end{equation*}
Observe that the Bayes factor in (\ref{BF:2F1}) can be handled using a one-dimensional integral. The Laplace approximation approach in \cite{liang:2008} may also be employed to evaluate the integral over the entire real line. Nevertheless, it seems difficult to choose the appropriate types of the approximations that we should employ in practice and to assess the quality of these approximations, especially when both $p$ and $r$ are extremely large. Of particular note here is that with the use of the above prior distributions, we can derive an analytical closed-form Bayes factor with an appropriate choice of $\beta$ in the following theorem. The proof is straightforward and is thus omitted for brevity.
\begin{theorem0} \label{thm:01}
With the priors given by (\ref{prior:1}) under $M_0$ and by (\ref{prior:2}), $(\ref{prior:3})$ and (\ref{beta:prime}) with $\kappa=r$ and $\beta = (n-p)/2-\alpha-2$ under $M_1$, the Bayes factor given by (\ref{BF:2F1}) turns out to be
\begin{equation}\label{BFequation}
BF_{10} = \frac{\Gamma{(p/2 + \alpha + 1/2)}\Gamma{((n - p )/2)}}{\Gamma{((n - 1)/2)}\Gamma{(\alpha + 1)}}\biggl(\frac{W_E}{W_T}\biggr)^{-(n - p - 2)/2 + \alpha}.
\end{equation}
\end{theorem0}

Notice that the Bayes factor in (\ref{BFequation}) has an explicit closed-form expression without integral representation, which can easily be calculated by using standard statistical software such as Matlab or R, and is readily accessible to non-statisticians in real applications. In other words, the Pearson type VI prior for the ratio of variance components provides a simple way of avoiding complex computational difficulties in the case where evaluation of the Bayes factor includes solving integrals. It is worth noting that the expression of Bayes factor given by (\ref{BFequation}) exactly coincides with the one in \cite{Maruyama:2009} for the balanced fixed effects model. Such an expression agreement is a consequence of the special choice $\kappa = r$ and $\beta = (n-p)/2-\alpha-2$ in the Pearson type VI distribution and may be unavailable for other choices of $\kappa$ and $\beta$, even if both fixed and random effects can be treated as random variables from the Bayesian viewpoint.

\begin{figure}[!htbp]
\begin{center}
\includegraphics[scale=0.7]{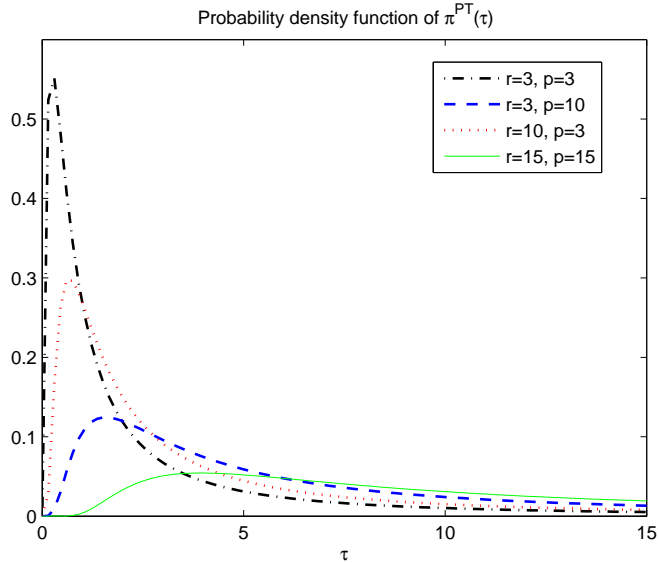}
\end{center}
\caption{The prior $\pi^{PT}(\tau)$ in (\ref{beta:prime}) with $\kappa=r$, $\alpha=-1/4$ and $\beta = (n-p)/2-\alpha-2$ for the different choices of ($r,~p$)}
\label{fig:01}
\end{figure}

At this stage, the hyperparameter $\alpha$ in the expression of Bayes factor (\ref{BFequation}) has not yet been assigned. It is well known that in Bayesian statistical analysis, choosing the hyperparameters of the prior distribution has a large impact on the behavior of Bayes factor. In this paper, we recommend $-1/2 \leq \alpha \leq 0$. It has been shown in the simulation studies that the proposed Bayes factor is quite robust to the choice of $\alpha \in [-1/2, ~0]$. Note that the prior $\pi^{PT}(\tau)$ with $\beta = (n-p)/2-\alpha-2$ depends on the sample size $n$; this kind of prior has also been adopted by many authors; see, for example, \cite{Maruyama:2009}, \cite{liang:2008}, \cite{Maru:Geor:2010}, to mention just a few.  As the sample size grows, the prior $\pi^{PT}(\tau)$ has a density in the right tail that behaves like $\tau^{-(\alpha+2)}$, leading to a very fat tail for small value of $\alpha$. Furthermore, it can be seen from Figure \ref{fig:01} that its mode also tends to 0 and thus this prior puts more weight to small values of $\tau$, an attractive property considered by \cite{Gust:Hoss:2006}. It should be mentioned that other optimal choices of these hyperparameters such as the one based on the empirical Bayes criterion can be further explored in future work.

\section{Model selection consistency} \label{section3}

From the Bayesian theoretical point of view, it is of particular interest to investigate the asymptotic behavior of Bayes factors such as consistency when the sample size approaches infinity. As mentioned in Section \ref{section1}, consistency means that the true model will be selected if enough data are provided, assuming that one of the competing models is true. This is formally introduced by \cite{Fern:2001} and later is called ``model selection consistency'' by \cite{liang:2008}.

In this section, we mainly focus on the consistency properties of the resulting Bayes factor in (\ref{BFequation}) for hypothesis testing in the sense that when the sample size approaches infinity, the Bayes factor goes to infinity when the alternative model $M_1$ is true, while it goes to 0 when the null model $M_0$ is true. We are now in a position to summarize the Bayes factor consistency under the three asymptotic scenarios described in Section \ref{section1} as follows.
\begin{theorem0}\label{theorem:2}
Consider the balanced one-way random effects ANOVA model (\ref{model1}) and the Bayes factor for testing $M_0$ against $M_1$ in (\ref{hpy:0}) with the priors given by ($\ref{prior:1}$), ($\ref{prior:2}$) and ($\ref{prior:3}$) as well as (\ref{beta:prime}) when $\kappa = r$ and $\beta = p(r-1)/2-\alpha-2$.
\begin{itemize}
\item [(a)] Under Scenario $1$, if $r$ goes to infinity, but $p \geq 2$ is fixed, then the Bayes factor in (\ref{BFequation}) is consistent whichever model is true.
\item [(b)] Under Scenario $2$, if $p$ goes to infinity, but $r \geq 2$ is fixed, then the Bayes factor in (\ref{BFequation}) is consistent under $M_0$ and under $M_1$ when $\sigma^2_a/\sigma^2 > h(r)$, while the Bayes factor is inconsistent when $\sigma^2_a/\sigma^2 < h(r)$, where
    \be \label{incons:region}
    h(r) = r^{1/(r-1)} - 1.
    \ee
\item [(c)] Under Scenario $3$, if both $p$ and $r$ go to infinity, then the Bayes factor in (\ref{BFequation}) is consistent whichever model is true.
\end{itemize}
\end{theorem0}
\begin{proof}
See Appendix for the proof.
\end{proof}

\begin{figure}[!htbp]
\begin{center}
\includegraphics[scale=0.7]{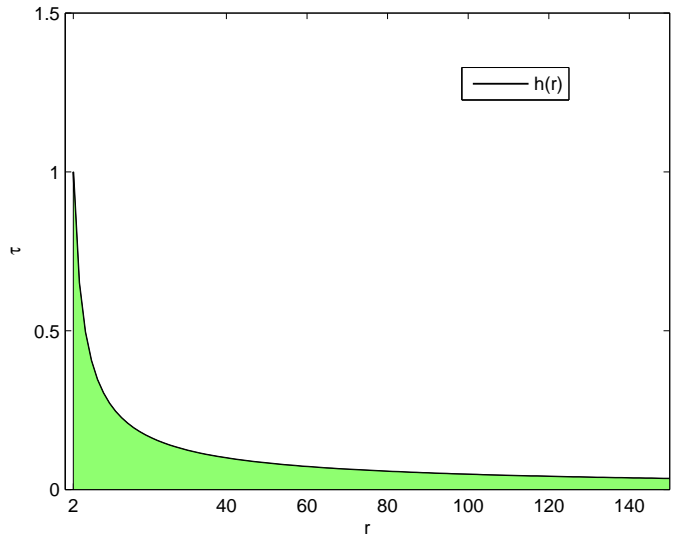}
\end{center}
\caption{The function h(r) in (\ref{incons:region}) used to determine the inconsistency region}
\label{fig:02}
\end{figure}

The above theorem has established the consistency properties of the proposed Bayes factor (\ref{BFequation}), which states convergence in probability of the true model asymptotically being chosen when the sample size approaches infinity under the three different asymptotic scenarios listed above. It should be noted that under Scenario 2 there exists an inconsistency region (the shaded area in Figure \ref{fig:02}) located in a small neighborhood of the null model. This inconsistency region can be characterized by the function $h(r)$ in (\ref{incons:region}), a decreasing convex function in $r$, satisfying $\lim_{r\go \infty} h(r)= 0$. See Figure \ref{fig:02}. Under the balanced fixed effects ANOVA model, \cite{Maruyama:2009} recently developed a new closed-form Bayes factor for testing whether the fixed effects are jointly significant and also derived a similar inconsistency region given by
\begin{equation*}
  \lim_{p\go \infty}\frac{\sum_{i}a_i^2}{p\sigma^2}~>~h(r).
\end{equation*}
Furthermore, \cite{Maruyama:2009} justified that the existence of the inconsistency region is quite reasonable under Scenario 2 from the predictive Bayesian viewpoint. It is of interest to note here that the two inconsistency regions are quite similar, which indicates that the variance of the random effects $\sigma_a^2$ in the random effects model plays a similar role as the limit of $\sum_ia_i^2/p$ in the fixed effects model.

\section{Numerical results} \label{section4}

In this section, we aim to numerically illustrate the finite sample performance of the Bayes factor in (\ref{lik:integral}) for the Pearson type VI prior with various choices of the hyperparameters $(\alpha,~\beta,~\kappa)$ through some simulation studies. For this end, we describe how the data sets in the balanced one-way random effects model (\ref{model1}) are generated. Under $M_0$, the samples are simulated with $\mu = 0$, $\sigma^2=1$ and $\sigma_a^2=0$, and under $M_1$, the samples are simulated with $\mu = 0$, $\sigma^2=1$ and $\sigma^2_a$ is taken to be one of the five different values $\{0.5, ~1, ~2, ~3, ~5\}$. For each case, we generate data sets with various values of $p$ and $r$ in order to mimic the three different kinds of asymptotic scenarios. We analyze $10,000$ simulated data sets for each case with various choices of $p$ and $r$. The decision criterion used in the simulation study is to select $M_1$ if the Bayes factor $BF_{10}>1$ and $M_0$ otherwise.

We firstly consider the performance of Bayes factor (\ref{BFequation}) summarized in Theorem \ref{theorem:2} when $\kappa=r$, $\beta = (n-p)/2-\alpha-2$ and values of $\alpha \in [-1/2,~0]$ used are $-1/2,~-1/4,~-1/5$ and $-1/10$. The relative frequency of choosing the true model under the three different scenarios is shown in Tables \ref{table:sim01}, \ref{table:sim02} and \ref{table:sim03}, respectively. Rather than providing exhaustive results based on these simulations, we merely highlight the most important findings from the first three tables here. (i) It can be concluded that the Bayes factor in (\ref{BFequation}) is fully consistent under the null hypothesis because the relative frequency of choosing the null model is consistently closer to 1 as the sample size becomes large. (ii) From the fourth column of Table \ref{table:sim02} associated with $\sigma_a^2=0.5$, it can be seen that the Bayes factor approaches 0 as the sample size increases. This phenomenon indicates that even though the model $M_1$ is true, the Bayes factor still chooses the null model with probability $1$ and thus fails to be asymptotically consistent. Such a conclusion exactly matches the statements in part (b) of Theorem \ref{theorem:2}, because in the simulation setup we have $\sigma_a^2/\sigma^2 = 0.5 < h(2) = 1$ when $r=2$. Similar conclusions can also be reached for $\sigma_a^2=1$, but the Bayes factor converges to 0 much slowly. The Bayes factor is fully consistent for $\sigma_a^2$ = $2, ~3$ and $5$ when the sample size becomes large because of $\sigma_a^2/\sigma^2 > h(2)$. (iii) As one would expect, the Bayes factors are fairly robust to the choice of the hyperparameter $\alpha \in[-1/2,~0]$ because similar results are obtained for the various values of $\alpha$ shown in the Tables \ref{table:sim01}, \ref{table:sim02} and \ref{table:sim03}. In conclusion, the simulation results clearly support the consistency claims made in Theorem \ref{table:sim02}. Furthermore, it is also noteworthy that the performance of the proposed Bayes factor is quite satisfactory even for the moderate values of $p$ and/or $r$.

We also investigate the performance of Bayes factor in (\ref{lik:integral}) under the Pearson type VI prior for various choices of the hyperparameters $(\alpha,~\beta,~\kappa)$, as suggested by one referee. Four different choices of these parameters are considered in the following simulation study. Choosing $(\alpha,~\beta,~\kappa) = (-1/2,~0,~1)$ results in the WG prior (\ref{hyp:gprior}) considered by \cite{Westfall:1996}. The choice of $(\alpha,~\beta,~\kappa) = (-1/2,~0,~1/n)$ yields the hyper-$g/n$ prior studied by \cite{liang:2008} for the hyperparameter $g$ in Zellner's $g$-prior. The choice of $(\alpha,~\beta,~\kappa) = (0,~-1/2,~1)$ leads to the prior suggested by \cite{Box:Tiao:1973} and \cite{Berg:Deel:1988}, and finally, $(\alpha,~\beta,~\kappa) = (0,~0,~r)$ corresponds to the prior derived by choosing a uniform prior on $[0,~1]$ for the parameter $r\tau/(1+r\tau)$. It should be mentioned that the numerical integration techniques have been employed to approximate the Bayes factors under these four different choices of $(\alpha,~\beta,~\kappa)$ because their expressions are not analytically tractable.

Tables \ref{table:sim04}, \ref{table:sim05} and \ref{table:sim06} summarize the results based on the above simulation setup. The following findings can be drawn from these simulation studies. (i) The Bayes factor with the four choices of $(\alpha,~\beta,~\kappa)$ generates compatible results in most cases, except for the case $(\alpha,~\beta,~\kappa)= (-1/2,~0,~1/n)$, which can be viewed as the most conservative criterion, having associated the smallest relative frequency of rejecting wrongly $M_0$. (ii) Of particular note is that there is no inconsistency region under Scenario 2, and thus we may conclude that under the three asymptotic scenarios, the Bayes factor in (\ref{lik:integral}) with these four different choices of $(\alpha,~\beta,~\kappa)$ is fully consistent whichever the true model is. (iii) As the sample size becomes large, the relative frequency of choosing the true model significantly increases and gets closer to each other under the different proposals; a similar conclusion can also be made when the variance $\sigma_a^2$ becomes large. Consequently, we may conclude that the Bayes factor in (\ref{lik:integral}) under the different priors for $\tau$ behaves very similarly.

As mentioned by a referee, it is noteworthy that the numerical results presented above only illustrated the finite sample performance of the proposed Bayes factor for the different hyperparameter values. We make no claim about the convergence rate of the Bayes factor from our simulation study. The convergence rate of the proposed Bayes factor under the different asymptotic scenarios also deserves further exploration.

In summary, one appealing advantage of the proposed Bayes factor in (\ref{BFequation}) is its explicit closed-form expression without integral representation. It has been observed from Table \ref{table:sim01} to Table \ref{table:sim06} that the behavior of the proposed Bayes factor is compatible with the one for other various choices of the hyperparameters $(\alpha,~\beta,~\kappa)$, except for the case under Scenario 2 due to the presence of a small inconsistency region around the null model.  Note that such an inconsistency region may be avoided with some other specific choices of the hyperparameters shown in the simulation study. However, a drawback of these choices is that the Bayes factor may not be analytically available.

\section{Concluding remarks} \label{section5}

In this paper, we have developed an explicit closed-form Bayes factor without integral representation, which can be easily calculated and is readily applicable to the problem of hypothesis testing under the balanced one-way random effects model. It is shown that under the different asymptotic scenarios described in Section \ref{section1}, the Bayes factor (\ref{BFequation}) is fully consistent under the null model, and is also consistent under the alternative model except for a small inconsistency region around the null model characterized by equation (\ref{incons:region}). Such the inconsistency region is the price we have to pay for deriving the closed-form of the Bayes factor, as mentioned by one referee. In addition, the referee also presumes that the inconsistency region may disappear when the prior is independent of the sample size. Such the presumption is quite understandable, but further investigation is needed, especially when the number of observation per unit goes to infinity. Looking at the simulation studies above, it seems that the inconsistency region may also disappear with some other specific choices of the hyperparameters $(\alpha,~\beta,~\kappa)$ in the Pearson type VI distribution, whereas the corresponding theoretical properties of Bayes factors with those specific choices under the three asymptotic scenarios are currently under investigation and will be reported elsewhere.

In some practical situations, unbalanced data may occur because of physical limitations and/or cost constraints. \cite{Min:Sun:2010} have recently generalized the results of \cite{Maruyama:2009} for the balanced fixed effects model to the ones for the unbalanced fixed effects model. In an ongoing project, it will be interesting for researchers to study the consistency of Bayes factor under the Pearson type VI prior for the unbalanced random effects model.

\vskip6mm
\noindent {\bf Acknowledgements.} \ \ The authors are very grateful to the Editor and anonymous referees for their constructive comments and suggestions that have substantially improved the appearance of this paper.

\section{Appendix} \label{section6}

Before proving Theorem \ref{theorem:2}, we first provide three useful lemmas. The proofs of these lemmas are straightforward based on several lemmas in \cite{GarcSun:2007} and are thus omitted here for simplicity.
\begin{lemma} \label{lemma:1}
Under Scenario $1$, for any fixed $p\geq2$, when the model $M_0$ is true,
\begin{equation}\label{scenario1:WEWT0}
\plim_{r\go \infty}\biggl( \frac{W_{E}}{W_{T}}\biggr)^{-(n-p-2)/2 + \alpha} = \exp\biggl(\frac{pW_H}{2(p-1)}\biggr),
\end{equation}
and when the model $M_1$ is true,
\begin{equation}\label{scenario1:WEWT1}
\plim_{r\go \infty}\frac{W_{E}}{W_{T}} = \biggl(1 + \frac{\sigma_a^2c_1}{p-1}\biggr)^{-1},
\end{equation}
where $c_1 = \lim_{r\go\infty}W_H/(\sigma^2 + r\sigma^2_a)$, which follows a chi-square distribution with $p-1$ degrees of freedom and is hence a distribution free of $r$.
\end{lemma}

\begin{lemma} \label{lemma:2}
Under Scenario $2$: for any fixed $r \geq 2$,
\begin{equation}\label{scenario2:WEWT}
\plim_{p\go\infty}\bigg\{ \frac{W_{H}}{W_{E}}\Bigl(\frac{n-p}{p-1}\Bigr)\bigg\}  =
\begin{cases}
1,  & \mbox{if the model } M_{0}\mbox{ is true,} \\[3pt]
1+r\sigma_a^2/\sigma^2, & \mbox{if the model } M_{1}\mbox{ is true.}
\end{cases}
\end{equation}
\end{lemma}

\begin{lemma} \label{lemma:3}
Under Scenario $3$: both $r$ and $p$ goes to infinity,
\begin{equation}\label{scenario3:WEWT}
\plim_{\substack{r\go\infty \\ p \go \infty}} \bigg\{\frac{W_E}{W_T}\Bigl(\frac{r}{r-1}\Bigr)\bigg\}  =
\begin{cases}
1,  & \mbox{if the model } M_{0}\mbox{ is true,} \\[2pt]
\bigl(1+\sigma_a^2/\sigma^2\bigr)^{-1}, & \mbox{if the model } M_{1}\mbox{ is true.}
\end{cases}
\end{equation}
\end{lemma}

\vskip10mm

\noindent{\bf Proof of Theorem \ref{theorem:2}:}

It is well known that when $x$ is sufficiently large, the Stirling's approximation to the gamma function is given by
\begin{equation*}
\Gamma(\gamma_1 x + \gamma_2) \approx \sqrt{2\pi}e^{-\gamma_1 x}(\gamma_1 x)^{\gamma_1 x + \gamma_2 - 1/2},
\end{equation*}
where $f (x) \approx g(x)$ means the limit of $f(x)/g(x)$ is one as $x$ approaches infinity.

(a) By using Lemma \ref{lemma:1}, it is easy to show that, under $M_0$,
\begin{align*}
BF_{10} &=\frac{\Gamma{(p/2 + \alpha + 1/2)}\Gamma{((n - p )/2)}}{\Gamma{((n - 1)/2)}\Gamma{(\alpha + 1)}}\biggl(\frac{W_E}{W_T}\biggr)^{-(n - p - 2)/2 + \alpha}\\
&\approx \frac{\Gamma(p/2 + \alpha + 1/2)}{\Gamma(\alpha+1)}\Big(\frac{n}{2}\Big)^{-(p-1)/2} \exp\biggl(\frac{pW_H}{2(p-1)}\biggr),
\end{align*}
which obviously goes to $0$ under $M_0$ as $r$ goes to infinity because the distribution of $W_H$ under $M_0$ is free of $r$. Under $M_1$,
\begin{align*}
BF_{10} &\approx \frac{\Gamma(p/2 + \alpha + 1/2)}{\Gamma(\alpha+1)}\Big(\frac{n}{2}\Big)^{-(p-1)/2} \biggl(1 + \frac{\sigma^2_a c_1}{p-1}\biggr)^{(n-p-2)/2 - \alpha}\\
&\approx \frac{\Gamma(p/2 + \alpha + 1/2)}{\Gamma(\alpha+1)}\Big(\frac{p}{2}\Big)^{-(p-1)/2}r^{-(p-1)/2}\bigg(1 +\frac{c_1}{p-1}\sigma^2_a\bigg)^{rp/2},
\end{align*}
which goes to infinity under $M_1$ for $\sigma_a^2 >0$ as $r$ goes to infinity.

(b) Similarly, by using Lemma \ref{lemma:2}, it is easy to show that, under $M_0$,
\begin{align*}
BF_{10} &\approx c_2(\alpha,r)\Big(\frac{p}{2}\Big)^{\alpha + 1/2}\biggl(\frac{r-1}{r^{r/(r-1)}}\biggr)^{p(r -1)/2}\biggl(1 + \frac{p-1}{n - p}\biggr)^{(n-p-2)/2 - \alpha}\\
&\approx c_2(\alpha,r)\Big(\frac{1}{2}\Big)^{\alpha+1/2}\biggl(1 + \frac{1}{r-1}\biggr)^{-1-\alpha}p^{\alpha + 1/2}r^{-p/2}\biggl(1 - \frac{1/r}{p}\biggr)^{p(r-1)/2}\\
&\approx c_2(\alpha,r)\Big(\frac{1}{2}\Big)^{\alpha+1/2}\biggl(1 + \frac{1}{r-1}\biggr)^{-1-\alpha}p^{\alpha + 1/2}r^{-p/2}\exp{\bigg(\frac{r-1}{2r} \bigg)},
\end{align*}
which goes to $0$ under $M_0$ as $p$ goes to infinity. Here $c_2(\alpha,r)$ is a constant independent of $p$ and is given by
\begin{align*}
c_2(\alpha,r) = \frac{\sqrt{2\pi}r}{\Gamma(\alpha+1/2)(r-1)^{1/2}}.
\end{align*}
Under $M_1$, it is easy to see that
\begin{align*}
BF_{10} &\approx c_2(\alpha,r)\Big(\frac{p}{2}\Big)^{\alpha + 1/2}\biggl(\frac{r-1}{r^{r/(r-1)}}\biggr)^{p(r -1)/2}\bigg(1 + \frac{(1 + r\sigma^2_a/\sigma^2)(p-1)}{p(r-1)}\bigg)^{(n-p-2)/2 - \alpha}\\
&\approx c_2(\alpha,r)\Big(\frac{p}{2}\Big)^{\alpha + 1/2}r^{-pr/2}(r-1)^{p(r-1)/2}\bigg(1 + \frac{1 + r\sigma^2_a/\sigma^2}{r-1}\frac{p-1}{p}\bigg)^{(n-p-2)/2 -\alpha}\\
&\approx c_2(\alpha,r)\Big(\frac{p}{2}\Big)^{\alpha+1/2}\biggl(\frac{r-1}{r(1 + \sigma_a^2/\sigma^2)}\biggr)^{1 +\alpha }\bigg(\frac{(1+\sigma^2_a/\sigma^2)^{r-1}}{r}\bigg)^{p/2}\bigg(1 - \frac{(1 + r\sigma^2_a/\sigma^2)}{p\big(r(1 + \sigma^2_a/\sigma^2)\big)}\bigg)^{(n-p)/2} \\
&\approx c_2(\alpha,r)\Big(\frac{p}{2}\Big)^{\alpha+1/2}\biggl(\frac{r-1}{r(1 + \sigma_a^2/\sigma^2)}\biggr)^{1 +\alpha }\bigg(\frac{(1+\sigma^2_a/\sigma^2)^{r-1}}{r}\bigg)^{p/2}\exp{\bigg(\frac{-(1 + r\sigma^2_a/\sigma^2)(r-1)}{2r(1 + \sigma^2_a/\sigma^2)}\bigg)},
\end{align*}
which goes to infinity under $M_1$ when
\begin{equation*}
 \frac{(1+\sigma^2_a/\sigma^2)^{r-1}}{r} >1,
\end{equation*}
indicating that $\sigma^2_a/\sigma^2 > r^{1/(r-1)} - 1 = h(r)$.

(c) Using Lemma \ref{lemma:3}, under $M_0$,
\begin{align*}
BF_{10}&\approx c_3(a)n^{a + 1/2}\Bigl(\frac{1}{r}\Bigr)^{a + p/2}\Bigl(1 - \frac{1}{r}\Bigr)^{(n- p-1)/2}\Bigl( \frac{r-1}{r}\Bigr)^{-(n- p-2)/2 + a}\\
&\approx c_3(a)p^{a + 1/2}\Bigl(\frac{1}{r}\Bigr)^{(p - 1)/2}\Bigl(1 - \frac{1}{r}\Bigl)^{a + 1/2},
\end{align*}
which clearly goes to $0$ under $M_0$ as both $r$ and $p$ go to infinity, where
\begin{equation*}
c_3(a) = \frac{\sqrt{2\pi}}{2^{a+1/2}\Gamma{(a + 1)}}.
\end{equation*}
Furthermore, when $M_1$ is true,
\begin{align*}
BF_{10}&\approx c_3(a)n^{a + 1/2}\Bigl(\frac{1}{r}\Bigr)^{a + p/2}\Bigl(1 - \frac{1}{r}\Bigr)^{(n-p-1)/2}\biggl(\frac{(r-1)/r}{1 + \sigma_a^2/ \sigma^2}\biggr)^{-(n - p-2)/2 + a} \\
&\approx c_3(a)p^{a +1/2} \Bigl(\frac{1}{r}\Bigr)^{(p - 1)/2}\Bigl(1 - \frac{1}{r}\Bigr)^{a + 1/2}\biggl(1 + \frac{\sigma_a^2}{ \sigma^2}\biggr)^{(n - p)/2 - (1+a)}\\
&\approx c_3(a)p^{a + 1/2}\Bigl(1 - \frac{1}{r}\Bigr)^{a + 1/2}r^{ - (1 + a)/(r-1) + 1/2}\bigg(\frac{1 + \sigma_a^2/ \sigma^2}{{r^{1/(r-1)}}}\bigg)^{(n - p)/2 - (1 + a )},
\end{align*}
which approaches infinity when $(1 +\sigma_a^2/\sigma^2)/r^{1/(r-1)} > 1$.  Namely, $\sigma_a^2 > 0$ when both $r$ and $p$ go to infinity. This completes the proof.

\newpage

\begin{table}[!htbp]
\centering % centering table
%\small\addtolength{\tabcolsep}{-0.3pt}
\begin{tabular}{c    c       c       c       c        c        c     c  }
\multicolumn{5}{r}{} \\ \hline \hline
          &          & \multicolumn{6}{c}{$\sigma_a^2$}       \\ \cline{3-8}
$(p, ~r)$ &$\alpha$  & $0$   &$0.5$  &  $1$  &  $2$   & $3$    & $5$   \\ \hline
\multirow{4}{*}{$(2,~5)$}
          & $-1/2$   & 0.894 & 0.358 & 0.476 & 0.594  & 0.662  & 0.733   \\
          & $-1/4$   & 0.860 & 0.400 & 0.521 & 0.630  & 0.694  & 0.759  \\
          & $-1/5$   & 0.855 & 0.408 & 0.528 & 0.637  & 0.700  & 0.762   \\
          & $-1/10$  & 0.845 & 0.421 & 0.540 & 0.647  & 0.707  & 0.770 \\ \hline
\multirow{4}{*}{$(2, ~10)$}
          & $-1/2$   & 0.937 & 0.436 & 0.562 & 0.677  & 0.729  & 0.790  \\
          & $-1/4$   & 0.914 & 0.470 & 0.593 & 0.700  & 0.751  & 0.807 \\
          & $-1/5$   & 0.910 & 0.475 & 0.597 & 0.704  & 0.755  & 0.810  \\
          & $-1/10$  & 0.901 & 0.483 & 0.606 & 0.711  & 0.766  & 0.816\\ \hline
\multirow{4}{*}{$(2, ~50)$}
          & $-1/2$   & 0.973 & 0.660 & 0.753 & 0.824  & 0.851  & 0.883  \\
          & $-1/4$   & 0.964 & 0.678 & 0.765 & 0.833  & 0.859  & 0.887 \\
          & $-1/5$   & 0.962 & 0.680 & 0.767 & 0.834  & 0.860  & 0.888  \\
          & $-1/10$  & 0.959 & 0.686 & 0.770 & 0.836  & 0.862  & 0.890\\ \hline
\multirow{4}{*}{$(2, ~100)$}
          & $-1/2$   & 0.984 & 0.734 & 0.812 & 0.866  & 0.891  & 0.914  \\
          & $-1/4$   & 0.979 & 0.747 & 0.820 & 0.871  & 0.897  & 0.917 \\
          & $-1/5$   & 0.978 & 0.748 & 0.821 & 0.872  & 0.897  & 0.918  \\
          & $-1/10$  & 0.976 & 0.752 & 0.823 & 0.874  & 0.899  & 0.919\\ \hline
\multirow{4}{*}{$(2, ~500)$}
          & $-1/2$   & 0.993 & 0.865 & 0.901 & 0.933  & 0.944  & 0.954  \\
          & $-1/4$   & 0.991 & 0.870 & 0.905 & 0.936  & 0.945  & 0.955\\
          & $-1/5$   & 0.991 & 0.871 & 0.905 & 0.936  & 0.946  & 0.956 \\
          & $-1/10$  & 0.990 & 0.872 & 0.906 & 0.937  & 0.956  & 0.967 \\ \hline \hline
\end{tabular}
\caption{The relative frequency of choosing the true model under the Bayes factor in (\ref{BFequation}) with four different values of the hyperparameter $\alpha$ for a fixed value of $p$ and increasing values of $r$ in the $10,000$ simulations.}
\label{table:sim01}
\end{table}

\begin{table}[!htbp]
\centering % centering table
%\small\addtolength{\tabcolsep}{-0.3pt}
\begin{tabular}{c    c       c       c       c        c        c     c  }
\multicolumn{5}{r}{} \\ \hline \hline
          &          & \multicolumn{6}{c}{$\sigma_a^2$}       \\ \cline{3-8}
$(p, ~r)$ &$\alpha$  & $0$   &$0.5$  &  $1$  &  $2$   & $3$    & $5$   \\ \hline
\multirow{4}{*}{$(5,~2)$}
          & $-1/2$   & 0.872 & 0.329 & 0.486 & 0.687  & 0.788  & 0.887   \\
          & $-1/4$   & 0.825 & 0.405 & 0.570 & 0.752  & 0.838  & 0.918  \\
          & $-1/5$   & 0.816 & 0.425 & 0.586 & 0.762  & 0.847  & 0.922   \\
          & $-1/10$  & 0.800 & 0.446 & 0.612 & 0.780  & 0.861  & 0.930 \\ \hline
\multirow{4}{*}{$(10, ~2)$}
          & $-1/2$   & 0.950 & 0.266 & 0.495 & 0.768  & 0.891  & 0.966  \\
          & $-1/4$   & 0.923 & 0.338 & 0.576 & 0.823  & 0.923  & 0.979 \\
          & $-1/5$   & 0.916 & 0.350 & 0.589 & 0.832  & 0.928  & 0.981  \\
          & $-1/10$  & 0.908 & 0.378 & 0.615 & 0.848  & 0.936  & 0.983\\ \hline
\multirow{4}{*}{$(50, ~2)$}
          & $-1/2$   & 1.000 & 0.080 & 0.500 & 0.961  & 0.998  & 1.000  \\
          & $-1/4$   & 1.000 & 0.108 & 0.561 & 0.973  & 0.999  & 1.000 \\
          & $-1/5$   & 1.000 & 0.112 & 0.572 & 0.979  & 0.999  & 1.000  \\
          & $-1/10$  & 1.000 & 0.123 & 0.592 & 0.979  & 1.000  & 1.000\\ \hline
\multirow{4}{*}{$(100, ~2)$}
          & $-1/2$   & 1.000 & 0.025 & 0.500 & 0.996  & 1.000  & 1.000  \\
          & $-1/4$   & 1.000 & 0.033 & 0.559 & 0.998  & 1.000  & 1.000 \\
          & $-1/5$   & 1.000 & 0.035 & 0.569 & 0.998  & 1.000  & 1.000  \\
          & $-1/10$  & 1.000 & 0.039 & 0.588 & 0.998  & 1.000  & 1.000\\ \hline
\multirow{4}{*}{$(500, ~2)$}
          & $-1/2$   & 1.000 & 0.000 & 0.494 & 1.000  & 1.000  & 1.000  \\
          & $-1/4$   & 1.000 & 0.000 & 0.528 & 1.000  & 1.000  & 1.000 \\
          & $-1/5$   & 1.000 & 0.000 & 0.534 & 1.000  & 1.000  & 1.000  \\
          & $-1/10$  & 1.000 & 0.000 & 0.544 & 1.000  & 1.000  & 1.000\\ \hline \hline
\end{tabular}
\caption{The relative frequency of choosing the true model under the Bayes factor in (\ref{BFequation}) with four different values of the hyperparameter $\alpha$ for a fixed value of $r$ and increasing values of $p$ in the $10,000$ simulations.}
\label{table:sim02}
\end{table}

\begin{table}[!htbp]
\centering % centering table
%\small\addtolength{\tabcolsep}{-0.3pt}
\begin{tabular}{c    c       c       c       c        c        c     c  }
\multicolumn{5}{r}{} \\ \hline \hline
          &          & \multicolumn{6}{c}{$\sigma_a^2$}       \\ \cline{3-8}
$(p, ~r)$ &$\alpha$  & $0$   &$0.5$  &  $1$  &  $2$   & $3$    & $5$   \\ \hline
\multirow{4}{*}{$(2,~2)$}
          & $-1/2$   & 0.764 & 0.348 & 0.423 & 0.520  & 0.580  & 0.654   \\
          & $-1/4$   & 0.714 & 0.407 & 0.485 & 0.533  & 0.634  & 0.701  \\
          & $-1/5$   & 0.705 & 0.417 & 0.496 & 0.587  & 0.642  & 0.708   \\
          & $-1/10$  & 0.690 & 0.434 & 0.511 & 0.600  & 0.657  & 0.721 \\ \hline
\multirow{4}{*}{$(10, ~5)$}
          & $-1/2$   & 0.996 & 0.465 & 0.812 & 0.967  & 0.990  & 0.998  \\
          & $-1/4$   & 0.994 & 0.519 & 0.841 & 0.973  & 0.993  & 0.998 \\
          & $-1/5$   & 0.993 & 0.527 & 0.845 & 0.975  & 0.993  & 0.998  \\
          & $-1/10$  & 0.993 & 0.542 & 0.855 & 0.977  & 0.993  & 0.999\\ \hline
\multirow{4}{*}{$(5, ~10)$}
          & $-1/2$   & 0.992 & 0.630 & 0.836 & 0.942  & 0.972  & 0.989  \\
          & $-1/4$   & 0.987 & 0.668 & 0.858 & 0.950  & 0.976  & 0.990 \\
          & $-1/5$   & 0.986 & 0.675 & 0.861 & 0.951  & 0.977  & 0.991  \\
          & $-1/10$  & 0.984 & 0.686 & 0.867 & 0.954  & 0.978  & 0.991\\ \hline
\multirow{4}{*}{$(10, ~10)$}
          & $-1/2$   & 1.000 & 0.748 & 0.953 & 0.995  & 0.991  & 1.000  \\
          & $-1/4$   & 1.000 & 0.778 & 0.960 & 0.997  & 0.993  & 1.000 \\
          & $-1/5$   & 1.000 & 0.783 & 0.962 & 0.997  & 0.993  & 1.000  \\
          & $-1/10$  & 1.000 & 0.793 & 0.964 & 0.997  & 0.993  & 1.000\\ \hline
\multirow{4}{*}{$(50, ~25)$}
          & $-1/2$   & 1.000 & 1.000 & 1.000 & 1.000  & 1.000 & 1.000  \\
          & $-1/4$   & 1.000 & 1.000 & 1.000 & 1.000  & 1.000  & 1.000 \\
          & $-1/5$   & 1.000 & 1.000 & 1.000 & 1.000  & 1.000  & 1.000  \\
          & $-1/10$  & 1.000 & 1.000 & 1.000 & 1.000  & 1.000  & 1.000\\ \hline
\multirow{4}{*}{$(25, ~50)$}
          & $-1/2$   & 1.000 & 1.000 & 1.000 & 1.000  & 1.000  & 1.000  \\
          & $-1/4$   & 1.000 & 1.000 & 1.000 & 1.000  & 1.000  & 1.000 \\
          & $-1/5$   & 1.000 & 1.000 & 1.000 & 1.000  & 1.000  & 1.000  \\
          & $-1/10$  & 1.000 & 1.000 & 1.000 & 1.000  & 1.000  & 1.000\\ \hline   \hline
\end{tabular}
\caption{The relative frequency of choosing the true model under the Bayes factor in (\ref{BFequation}) with four different values of the hyperparameter $\alpha$ for increasing values of $p$ and $r$ in the $10,000$ simulations.}
\label{table:sim03}
\end{table}

\begin{table}[!htbp]
\centering % centering table
%\small\addtolength{\tabcolsep}{-0.3pt}
\begin{tabular}{c    c    c c    c       c       c        c        c     c  }
\multicolumn{5}{r}{} \\ \hline \hline
          &         &        &          &\multicolumn{6}{c}{$\sigma_a^2$}       \\ \cline{5-10}
$(p, ~r)$ &$\alpha$ &$\beta$ & $\gamma$ & $0$   &$0.5$  &  $1$  &  $2$   & $3$    & $5$   \\ \hline
\multirow{4}{*}{$(2,~5)$}
          & $-1/2$  &   0    &    1     & 0.853 & 0.411 & 0.530 & 0.643  & 0.702  & 0.764  \\
          & $-1/2$  &   0    &  $1/n$   & 0.951 & 0.250 & 0.370 & 0.505  & 0.580  & 0.662  \\
          &   0     & $-1/2$ &    1     & 0.899 & 0.350 & 0.469 & 0.586  & 0.656  & 0.729  \\
          &   0     &   0    &   $r$    & 0.797 & 0.479 & 0.585 & 0.690  & 0.741  & 0.797  \\ \hline
\multirow{4}{*}{$(2, ~10)$}
          & $-1/2$  &   0    &    1     & 0.893 & 0.492 & 0.617 & 0.718  & 0.768  & 0.821  \\
          & $-1/2$  &   0    &  $1/n$   & 0.979 & 0.323 & 0.460 & 0.592  & 0.662  & 0.734  \\
          &   0     & $-1/2$ &    1     & 0.932 & 0.444 & 0.570 & 0.681  & 0.733  & 0.795  \\
          &   0     &   0    &   $r$    & 0.813 & 0.582 & 0.683 & 0.769  & 0.811  & 0.855  \\ \hline
\multirow{4}{*}{$(2, ~50)$}
          & $-1/2$  &   0    &    1     & 0.945 & 0.707 & 0.787 & 0.847  & 0.871  & 0.897   \\
          & $-1/2$  &   0    &  $1/n$   & 0.996 & 0.567 & 0.680 & 0.766  & 0.809  & 0.849   \\
          &   0     & $-1/2$ &    1     & 0.964 & 0.677 & 0.765 & 0.832  & 0.858  & 0.887   \\
          &   0     &   0    &   $r$    & 0.819 & 0.793 & 0.852 & 0.896  & 0.901  & 0.927   \\ \hline
\multirow{4}{*}{$(2, ~100)$}
          & $-1/2$  &   0    &    1     & 0.964 & 0.767 & 0.834 & 0.884  & 0.906  & 0.925   \\
          & $-1/2$  &   0    &  $1/n$   & 0.999 & 0.655 & 0.751 & 0.823  & 0.856  & 0.887   \\
          &   0     & $-1/2$ &    1     & 0.978 & 0.749 & 0.821 & 0.872  & 0.898  & 0.918   \\
          &   0     &   0    &   $r$    & 0.830 & 0.847 & 0.889 & 0.926  & 0.938  & 0.947   \\ \hline
\multirow{4}{*}{$(2, ~500)$}
          & $-1/2$  &   0    &    1     & 0.993 & 1.000 & 1.000 & 1.000  & 1.000  & 1.000  \\
          & $-1/2$  &   0    &  $1/n$   & 0.991 & 1.000 & 1.000 & 1.000  & 1.000  & 1.000  \\
          &   0     & $-1/2$ &    1     & 0.991 & 1.000 & 1.000 & 1.000  & 1.000  & 1.000  \\
          &   0     &   0    &   $r$    & 0.990 & 1.000 & 1.000 & 1.000  & 1.000  & 1.000  \\ \hline \hline
\end{tabular}
\caption{The relative frequency of choosing the true model under the Bayes factor in (\ref{lik:integral}) for the Pearson type VI prior with different values of the hyperparameters ($\alpha,~\beta, ~\kappa$) for a fixed value of $p$ and increasing values of $r$ in the $10,000$ simulations.}
\label{table:sim04}
\end{table}

\begin{table}[!htbp]
\centering % centering table
%\small\addtolength{\tabcolsep}{-0.3pt}
\begin{tabular}{c    c    c c    c       c       c        c        c     c  }
\multicolumn{5}{r}{} \\ \hline \hline
          &         &        &          &\multicolumn{6}{c}{$\sigma_a^2$}       \\ \cline{5-10}
$(p, ~r)$ &$\alpha$ &$\beta$ & $\gamma$ & $0$   &$0.5$  &  $1$  &  $2$   & $3$    & $5$   \\ \hline
\multirow{4}{*}{$(5,~2)$}
          & $-1/2$  &   0    &    1     & 0.801 & 0.445 & 0.611 & 0.780  & 0.860  & 0.930   \\
          & $-1/2$  &   0    &  $1/n$   & 0.945 & 0.171 & 0.293 & 0.489  & 0.627  & 0.776   \\
          &   0     & $-1/2$ &    1     & 0.870 & 0.332 & 0.491 & 0.692  & 0.791  & 0.889   \\
          &   0     &   0    &   $r$    & 0.763 & 0.499 & 0.659 & 0.812  & 0.886  & 0.943   \\ \hline
\multirow{4}{*}{$(10, ~2)$}
          & $-1/2$  &   0    &    1     & 0.850 & 0.497 & 0.722 & 0.908  & 0.963  & 0.991   \\
          & $-1/2$  &   0    &  $1/n$   & 0.982 & 0.115 & 0.290 & 0.584  & 0.756  & 0.913   \\
          &   0     & $-1/2$ &    1     & 0.908 & 0.555 & 0.614 & 0.928  & 0.972  & 0.994   \\
          &   0     &   0    &   $r$    & 0.809 & 0.562 & 0.776 & 0.930  & 0.973  & 0.995   \\ \hline
\multirow{4}{*}{$(50, ~2)$}
          & $-1/2$  &   0    &    1     & 0.931 & 0.818 & 0.989 & 1.000  & 1.000  & 1.000   \\
          & $-1/2$  &   0    &  $1/n$   & 1.000 & 0.186 & 0.704 & 0.990  & 1.000  & 1.000   \\
          &   0     & $-1/2$ &    1     & 0.962 & 0.720 & 0.979 & 1.000  & 1.000  & 1.000   \\
          &   0     &   0    &   $r$    & 0.899 & 0.876 & 0.995 & 1.000  & 1.000  & 1.000   \\ \hline
\multirow{4}{*}{$(100, ~2)$}
          & $-1/2$  &   0    &    1     & 0.953 & 0.960 & 1.000 & 1.000  & 1.000  & 1.000   \\
          & $-1/2$  &   0    &  $1/n$   & 1.000 & 0.420 & 0.964 & 1.000  & 1.000  & 1.000   \\
          &   0     & $-1/2$ &    1     & 0.977 & 0.924 & 1.000 & 1.000  & 1.000  & 1.000   \\
          &   0     &   0    &   $r$    & 0.922 & 0.977 & 1.000 & 1.000  & 1.000  & 1.000   \\ \hline
\multirow{4}{*}{$(500, ~2)$}
          & $-1/2$  &   0    &    1     & 0.995 & 1.000 & 1.000 & 1.000  & 1.000  & 1.000  \\
          & $-1/2$  &   0    &  $1/n$   & 1.000 & 1.000 & 1.000 & 1.000  & 1.000  & 1.000  \\
          &   0     & $-1/2$ &    1     & 0.993 & 1.000 & 1.000 & 1.000  & 1.000  & 1.000  \\
          &   0     &   0    &   $r$    & 0.999 & 1.000 & 1.000 & 1.000  & 1.000  & 1.000  \\ \hline \hline
\end{tabular}
\caption{The relative frequency of choosing the true model under the Bayes factor in (\ref{lik:integral}) for the Pearson type VI prior with different values of the hyperparameters ($\alpha,~\beta, ~\kappa$) for a fixed value of $r$ and increasing values of $p$ in the $10,000$ simulations.}
\label{table:sim05}
\end{table}

\begin{table}[!htbp]
\centering % centering table
%\small\addtolength{\tabcolsep}{-0.3pt}
\begin{tabular}{c    c    c c    c       c       c        c        c     c  }
\multicolumn{5}{r}{} \\ \hline \hline
          &         &        &          &\multicolumn{6}{c}{$\sigma_a^2$}       \\ \cline{5-10}
$(p, ~r)$ &$\alpha$ &$\beta$ & $\gamma$ & $0$   &$0.5$  &  $1$  &  $2$   & $3$    & $5$   \\ \hline
\multirow{4}{*}{$(2,~2)$}
          & $-1/2$  &   0    &    1     & 0.734 & 0.382 & 0.461 & 0.553  & 0.613  & 0.683   \\
          & $-1/2$  &   0    &  $1/n$   & 0.842 & 0.255 & 0.320 & 0.419  & 0.483  & 0.564   \\
          &   0     & $-1/2$ &    1     & 0.799 & 0.308 & 0.381 & 0.478  & 0.539  & 0.619   \\
          &   0     &   0    &   $r$    & 0.713 & 0.407 & 0.486 & 0.578  & 0.634  & 0.702   \\ \hline
\multirow{4}{*}{$(10, ~5)$}
          & $-1/2$  &   0    &    1     & 0.938 & 0.805 & 0.959 & 0.993  & 0.998  & 1.000   \\
          & $-1/2$  &   0    &  $1/n$   & 0.998 & 0.376 & 0.746 & 0.951  & 0.985  & 0.997   \\
          &   0     & $-1/2$ &    1     & 0.966 & 0.734 & 0.936 & 0.992  & 0.997  & 1.000   \\
          &   0     &   0    &   $r$    & 0.859 & 0.890 & 0.980 & 0.997  & 0.999  & 1.000   \\ \hline
\multirow{4}{*}{$(5, ~10)$}
          & $-1/2$  &   0    &    1     & 0.945 & 0.795 & 0.916 & 0.974  & 0.988  & 0.996   \\
          & $-1/2$  &   0    &  $1/n$   & 0.999 & 0.485 & 0.748 & 0.903  & 0.951  & 0.980   \\
          &   0     & $-1/2$ &    1     & 0.969 & 0.746 & 0.856 & 0.965  & 0.984  & 0.993   \\
          &   0     &   0    &   $r$    & 0.844 & 0.881 & 0.956 & 0.983  & 0.994  & 0.998   \\ \hline
\multirow{4}{*}{$(10, ~10)$}
          & $-1/2$  &   0    &    1     & 0.968 & 0.952 & 0.994 & 0.999  & 1.000  & 1.000   \\
          & $-1/2$  &   0    &  $1/n$   & 1.000 & 0.731 & 0.948 & 0.994  & 1.000  & 1.000   \\
          &   0     & $-1/2$ &    1     & 0.984 & 0.930 & 0.991 & 0.991  & 1.000  & 1.000   \\
          &   0     &   0    &   $r$    & 0.872 & 0.981 & 0.998 & 1.000  & 1.000  & 1.000   \\ \hline
\multirow{4}{*}{$(25, ~50)$}
          & $-1/2$  &   0    &    1     & 0.997 & 1.000 & 1.000 & 1.000  & 1.000  & 1.000  \\
          & $-1/2$  &   0    &  $1/n$   & 1.000 & 1.000 & 1.000 & 1.000  & 1.000  & 1.000  \\
          &   0     & $-1/2$ &    1     & 0.990 & 1.000 & 1.000 & 1.000  & 1.000  & 1.000  \\
          &   0     &   0    &   $r$    & 0.911 & 1.000 & 1.000 & 1.000  & 1.000  & 1.000  \\ \hline
\multirow{4}{*}{$(50, ~25)$}
          & $-1/2$  &   0    &    1     & 0.996 & 1.000 & 1.000 & 1.000  & 1.000  & 1.000  \\
          & $-1/2$  &   0    &  $1/n$   & 1.000 & 1.000 & 1.000 & 1.000  & 1.000  & 1.000  \\
          &   0     & $-1/2$ &    1     & 0.998 & 1.000 & 1.000 & 1.000  & 1.000  & 1.000  \\
          &   0     &   0    &   $r$    & 0.933 & 1.000 & 1.000 & 1.000  & 1.000  & 1.000  \\ \hline \hline
\end{tabular}
\caption{The relative frequency of choosing the true model under the Bayes factor in (\ref{lik:integral}) for the Pearson type VI prior with different values of the hyperparameters ($\alpha,~\beta, ~\kappa$) for increasing values of $p$ and $r$ in the $10,000$ simulations.}
\label{table:sim06}
\end{table}

\bibliographystyle{annals}
\end{document}